\documentclass[prl,showpacs,aps,twocolumn]{revtex4}%
\usepackage{amsfonts}%
\usepackage{amsmath}%
\usepackage{amssymb}%
\usepackage{graphicx}

\begin{document}
\preprint{ }

\bigskip
\title{Exceptional points in the scattering continuum}

\author
{J. Oko{\l}owicz$^{\P}$, and M. P{\l}oszajczak$^{\$}$}

\affiliation{
$^{\P}$ Institute of Nuclear Physics, Polish Academy of Sciences, Radzikowskiego 152, PL-31342 Krak\'ow, Poland\\
$^{\$}$ Grand Acc\'{e}l\'{e}rateur National d'Ions Lourds (GANIL),
CEA/DSM -- CNRS/IN2P3, BP 5027, F-14076 Caen Cedex 05, France}

\date{\today}

\begin{abstract}
The manifestation of exceptional points in the scattering continuum of atomic nucleus is 
studied using the real-energy continuum shell model. It is shown that low-energy 
exceptional points appear for realistic values of coupling to the continuum and, hence, 
could be accessible experimentally. Experimental signatures are proposed which include 
the jump by $2\pi$ of the elastic scattering phase shift and a salient energy dependence of cross-sections in the vicinity of the exceptional point. 

\end{abstract}

\pacs{PACS number(s): 25.70.Ef, 03.65Vf, 21.60.Cs, 25.40.Cm, 25.40.Ep}

\bigskip

\maketitle

\section{Introduction}

The structure of loosely bound and unbound nuclei is strongly impacted by many-body 
correlations and non-perturbative coupling to the external environment of scattering 
states and decay channels \cite{Oko03,Mic09}. This is particularly important in 
exotic nuclei where new phenomena, at the borderline of nuclear structure and nuclear 
reactions, are expected. Some of them, like the halos \cite{Rii00}, the segregation of 
time scales  in the context of non-Hermitian Hamiltonians \cite{Kle85}, the alignment of near-threshold states with decay channels \cite{Ike68}, and the resonance crossings \cite{Zir83,Hei91} appear in various {\em open}  mesoscopic systems. Their universality is the consequence of the non-Hermitian nature of an eigenvalue problem in open quantum systems. 

Resonances are commonly found in quantum systems, independently of their interactions, 
building blocks and energy scales involved. Much interest is concentrated on resonance 
degeneracies, the so-called exceptional points (EPs) \cite{Zir83}. Their connection to avoided 
crossings and spectral properties of Hermitian systems \cite{Hei91a,Duk08} as well as the associated 
geometric phases have been discussed in simple models in considerable detail \cite{Hei98}. 
The interesting question is their manifestation in nuclear scattering experiments. Here, a much studied case was the $2^+$ doublet in $^8$Be \cite{Mar65,Pau66,Bar66,Bro66,Hin78}. Based on this example, von Brentano \cite{Bre90} discussed the width attraction for mixed resonances, and Hernand\'{e}z and Mondrag\'{o}n \cite{Her94} showed that the true crossing of resonances can be obtained by the variation of  two parameters in the Jordan block of rank two. In this latter analysis, it was shown that the resonating part of the scattering matrix (S-matrix) for one open channel and two internal states is compatible with the two-level formula of the R-matrix theory used in the experimental analysis of excitation functions of elastic scattering $^4$He($\alpha,\alpha_0$)$^4$He \cite{Hin78} and, hence, the $2^+$ doublet in $^8$Be may actually be close to the true resonance degeneracy.

Properties of atomic nucleus around the continuum threshold change rapidly with the 
nucleon number, the excitation energy and the coupling to the environment of scattering states. 
A consistent description of the interplay between scattering and resonant states requires 
an open system formulation of the nuclear shell model (see \cite{Oko03,Mic09,Zel06} 
for recent reviews).  The real-energy continuum shell model \cite{Fes58,Fan61,SMEC} 
provides a suitable unified framework with the help of an effective non-Hermitian 
Hamiltonian. In this work, for the first time we focus on a realistic model of an unbound atomic nucleus to see whether one or more EPs can appear in the low energy continuum for sensible parameters of the open quantum system Hamiltonian. In particular, we discuss possible experimental signatures of the EPs and show the evolution of these signatures in the vicinity of the EP. Finally, on the example of spectroscopic factors we demonstrate the entanglement of  resonance wave functions close to the EP.

\section{Formulation of the Continuum Shell Model}

Let us briefly review the Shell Model Embedded in the Continuum 
(SMEC) \cite{SMEC}, which is a recent realization of the real-energy continuum shell 
model. The total function space of an $A-$particle system consists of the set of 
square-integrable functions ${\cal Q}\equiv \{\psi_i^{A}\}$, used in the standard 
nuclear Shell Model (SM),  and the set of embedding scattering states 
${\cal P}\equiv \{\zeta_E^{c}\}$. These two sets are obtained by solving the Schr\"odinger equation, separately for discrete (SM) states (the closed quantum system) and for scattering states 
(the environment). Decay channels '$c$' are determined by the motion of an unbound 
particle in a state $l_j$ relative to the  $A-1$ nucleus with all nucleons on bounded 
single-particle (s.p.) orbits in the SM eigenstate $\psi_j^{A-1}$.  Using these function sets, 
one defines projection operators: 
\begin{eqnarray}
{\hat Q}=\sum_{i=1}^N|\psi_i^A\rangle\langle\psi_i^A|~;~~~~ {\hat P}=\int_0^\infty dE|\zeta_E\rangle\langle\zeta_E| \nonumber
\end{eqnarray}
and projected Hamiltonians: 
${\hat Q}H{\hat Q}\equiv H_{QQ}$, ${\hat P}H{\hat P}\equiv H_{PP}$, 
${\hat Q}H{\hat P}\equiv H_{QP}$, ${\hat P}H{\hat Q}\equiv H_{PQ}$.  
Assuming ${\cal Q}+{\cal  P}={\cal I}$, one can determine the third set of functions 
$\{\omega_i^{(+)}\}$ which contains the continuation of any SM eigenfunction 
$\psi_i^A$ in ${\cal P}$, and then construct the complete solution in ${\cal Q}+{\cal P}$ \cite{Oko03}. Recently, this approach has been extended to describe the two-proton radioactivity with the two-particle continuum \cite{Rot06}.

Open quantum system solutions in ${\cal Q}$, which include couplings to the 
environment of scattering states and decay channels, are obtained by solving 
the eigenvalue problem for the energy-dependent effective Hamiltonian:
\begin{eqnarray} 
{\cal H}_{QQ}(E)=H_{QQ}+H_{QP}G_P^{(+)}(E)H_{PQ} \ , 
\nonumber
\end{eqnarray}
where $H_{QQ}$ is the closed system Hamiltonian, $G_P^{(+)}(E)$ is the Green 
function for the motion of a single nucleon  in ${\cal P}$ subspace and $E$ is the 
energy of this nucleon (the scattering energy). Index '+' in $G_P^{(+)}$ stands for the
outgoing boundary in the scattering problem. ${\cal H}_{QQ}$ is non-Hermitian 
for unbound states and its eigenstates $|\Phi_\alpha\rangle$ are linear 
combinations of SM eigenstates $|\psi_i\rangle$. The eigenstates of ${\cal H}_{QQ}$ 
are biorthogonal; the left $|\Phi_\alpha\rangle$ and right $|\Phi_{\bar \alpha}\rangle$ 
eigenstates have the wave functions related by the complex conjugation. 
The orthonormality condition in the biorthogonal basis reads: 
$\langle\Phi_{\bar \alpha}|\Phi_{\beta}\rangle = \delta_{\alpha,\beta}$. Similarly, 
the matrix element of an operator ${\hat O}$ is 
$O_{\alpha\beta}=\langle\Phi_{\bar \alpha}|{\hat O}|\Phi_{\alpha}\rangle$. 

The scattering function $\Psi^c_E$ is a solution of a Schr\"{o}dinger equation 
in the total function space:
\begin{eqnarray}
\Psi^c_E=\zeta_E^c+\sum_{\alpha}a_{\alpha}{\tilde \Phi}_{\alpha} \ ,
\nonumber
\end{eqnarray}
where 
\begin{eqnarray}
a_{\alpha}\equiv\langle\Phi_{\alpha}|H_{QP}|\zeta^c_E\rangle/(E-{\cal E}_{\alpha}) \ ,
\nonumber
\end{eqnarray} 
and 
\begin{eqnarray}
{\tilde \Phi}_{\alpha}\equiv(1+G_P^{(+)}H_{PQ})\Phi_{\alpha} \ .
\nonumber
\end{eqnarray}
Inside of an interaction region, the dominant contributions to $\Psi^c_E$ 
are given by eigenfunctions $\Phi_{\alpha}$ of the effective non-Hermitian Hamiltonian \cite{Oko03}: 
\begin{eqnarray}
\Psi^c_E\sim\sum_{\alpha}a_{\alpha}\Phi_{\alpha} \ . 
\nonumber
\end{eqnarray}
For bounds states, eigenvalues ${\cal E}_\alpha(E)$ of ${\cal H}_{QQ}(E)$ are real and 
${\cal E}_{\alpha}(E)=E$. For unbound states, physical resonances can be identified 
with the narrow poles of the S-matrix \cite{Sie39,Mic09}, or using the Breit-Wigner approach which leads to a fixed-point condition \cite{Oko03,Zel06,Mad05}: 
\begin{eqnarray}
E_{\alpha}={\rm Re}\left( {\cal E}_{\alpha}(E) \right)|_{E=E_{\alpha}} ~;~~
\mathit{\Gamma}_{\alpha}=-2\,{\rm Im}\left( {\cal E}_{\alpha}(E) \right)|_{E=E_{\alpha}} 
\label{eq1}
\end{eqnarray}
Here it is assumed that the origin of ${\rm Re}\left( {\cal E} \right)$ is fixed at the lowest particle emission threshold.

An EP is a generic phenomenon in Hamiltonian systems. In our case, 
the EP can appear as a result of the continuum-coupling term 
$H_{QP}G_P^{(+)}(E)H_{PQ}$ for energies above the first particle emission 
threshold ($E>0$). The eigenvalue degeneracies are indicated by common roots 
of the two equations \cite{Zir83}:
\begin{eqnarray}
 \frac{\partial^{(\nu)}}{\partial {\cal E}} {\rm det}\left[{\cal H}_{QQ}\left(E;V_0\right)  -{\cal E}I\right] = 0~~~~~~~~~~\nu=0,1
\label{discr}
\end{eqnarray}
Single-root solutions of Eq. (2) correspond to EPs associated with decaying states. 
The maximal number of those roots is  $M_{max}=n(n-1)$, where $n$ is the number 
of states of given angular momentum $J$ and parity $\pi$. 
In quantum integrable models with at least two parameter-dependent 
integrals of motion one finds also double-root solutions which correspond to 
non-singular crossing of two levels with two different wave functions. Hence, 
the actual number of EPs in these systems is always smaller than $M_{max}$  \cite{Duk08}. 

The position of EPs in the spectrum of eigenvalues of ${\cal H}_{QQ}$ depends both 
on the chosen interaction and the energy $E$ of the system.
In general, eigenvalues of the energy-dependent effective Hamiltonian ${\cal H}_{QQ}(E)$ need not satisfy the fixed-point condition (\ref{eq1}) and hence need not correspond to poles of the 
S-matrix (resonances). In the following, we shall consider uniquely the case where
EPs are {\em identical} with double-poles of the S-matrix.

\section{Exceptional points in the scattering continuum of $^{16}{\rm Ne}$}

Let us investigate properties of EPs on the example of $^{16}$Ne. SM eigenstates in this nucleus correspond to a complicated mixture of configurations associated with the dynamics of the $^{16}$O core.  Our goal is to see if EPs can be possibly found in the scattering continuum of atomic nucleus  
at low excitation energies and for physical strength of the continuum coupling.
SMEC calculations are performed in $p_{1/2}, d_{5/2}, s_{1/2}$ model space. For $H_{QQ}$ we take the ZBM Hamiltonian \cite{ZBM} which correctly describes the configuration mixing around $N=Z=8$ shell closure. The residual coupling $H_{QP}$ between ${\cal Q}$ and the 
embedding continuum ${\cal P}$ is generated by the contact force: 
$H_{QP}=H_{PQ}=V_0\delta(r_1-r_2)$.  
For each $J^{\pi}$, the SM states $|\psi_i(J^{\pi})\rangle$ of the closed quantum system 
are interconnected via the coupling to common decay channels 
$[^{15}{\rm F}(K^{\pi})\otimes {\rm p}_{l_j}]_{E'}^{J^{\pi}}$ with 
$K^{\pi}=1/2^+, 5/2^+$, and $1/2^-$ which have the thresholds at $E=0$ 
(the elastic channel), 0.67 MeV, and 2.26 MeV, respectively. In the ZBM model space, 
these are all possible one-proton (1p) decay channels in $^{16}$Ne.

The size of a non-Hermitian correction to $H_{QQ}$ depends on two real parameters: 
the strength $V_0$ of the continuum coupling in $H_{QP}$ ($H_{PQ}$) and the system 
energy $E$. The range of relevant $V_0$ values can be determined, for example,  
by fitting decay widths of the lowest states in $^{15}$F.  For the present Hamiltonian, 
experimental decay widths of the ground state $1/2_1^+$   and the first excited state 
$5/2_1^+$ in $^{15}$F are reproduced using $V_0=-3500\pm 450$ MeV$\cdot$fm$^3$ 
and $V_0=-1100\pm 50$ MeV$\cdot$fm$^3$, respectively. The error bars in $V_0$ 
reflect experimental uncertainties of those widths. The weak dependence of 1p~decay widths on the sign of $V_0$ is generated by the channel-channel coupling and disappears in a single-channel case.

\begin{figure}[hbt]
\begin{center}
{\includegraphics[width=8cm,angle=00]{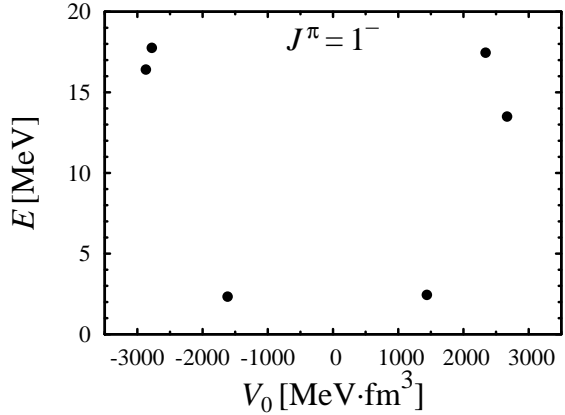}
\caption{The map of $J^{\pi}=1^-$ exceptional points in the continuum of $^{16}$Ne 
as found in SMEC. For more details, see the description in the text.}}
\label{fig1}
\end{center}
\end{figure}
Fig. 1 shows energies $E$ and strengths $V_0$ which correspond to $J^{\pi}=1^-$ 
EPs in the scattering continuum of  $^{16}$Ne.
Decay channels $[^{15}{\rm F}(K^{\pi})\otimes {\rm p}_{l_j}]_{E'}^{1^-}$ with 
$K^{\pi}=1/2^+, 5/2^+$, and $1/2^-$ have been included with proton partial waves:
$p_{1/2}, p_{3/2}$ for $K^{\pi}=1/2^+$, $p_{3/2}, f_{5/2}, f_{7/2}$ for  $K^{\pi}=5/2^+$, and 
$s_{1/2}, d_{3/2}$ for $K^{\pi}=1/2^-$. The number of $1^-$ SM states is 3 and, hence, 
the maximal number of $1^-$ EPs in SMEC could be 6. Indeed, all of them exist at  $E<20$ MeV
in a physical range of $V_0$ values (1100 MeV$\cdot$fm$^3<|V_0|<3500$ MeV$\cdot$fm$^3$).
They have been found by scanning the energy dependence of all eigenvalues over a certain range of $V_0$, searching for all real-energy crossings or width crossings (avoided crossings). Once found, we have tuned $V_0$ to find out whether these crossings evolve into EPs at some combination of $V_0$ and $E$. One should stress that the passage through EP always occurs if, e.g., 
the real-energy crossing moves towards $E=0$. Since such a crossing cannot move 
into the region $E<0$, therefore it converts into an avoided crossing via the formation of an EP.

The lowest EP in Fig. 1 is seen at $V_0^{(\rm cr)}=-1617.4$ MeV$\cdot$fm$^3$ 
and $E=2.33$ MeV. This EP corresponds to a degeneracy of the 
first two $1^-$ eigenvalues of ${\cal H}_{QQ}$ for $V_0<0$. 
\begin{figure}[hbt]
\begin{center}
\includegraphics[width=6cm,angle=00]{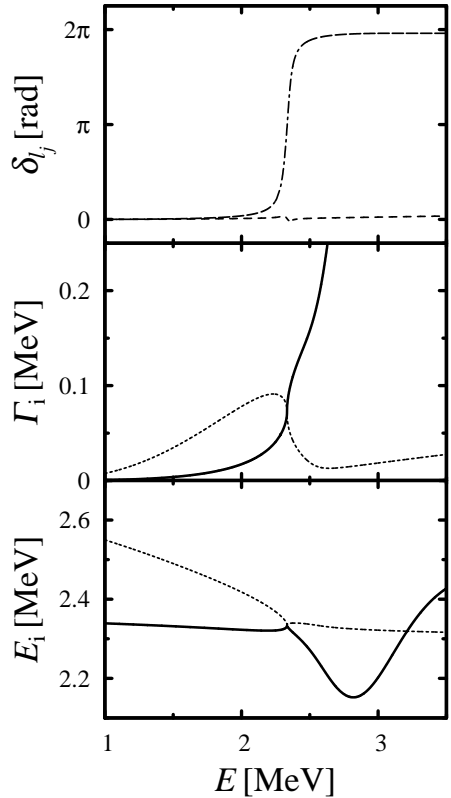}
\caption{The upper plot exhibits the elastic scattering phase shifts $\delta_{p_{1/2}}$ 
(dashed-dotted line) and $\delta_{p_{3/2}}$ (dashed line) for ${\rm p +} ^{15}{\rm F}$ 
reaction in $1^-$ partial waves at around the EP (the double-pole of the S-matrix) with $J^{\pi}=1^-$. Lower plots show real and imaginary parts of $1_1^-$ (solid line) and $1_2^-$ (dotted line) eigenvalues of the effective Hamiltonian ${\cal H}_{QQ}(E)$ as a function of the scattering energy $E$. For other details, see the description in the text.}
\label{fig2}
\end{center}
\end{figure}
Energy $E_i$ and width $\mathit{\Gamma}_i$ of $1^-_1$ and $1^-_2$ eigenvalues 
are shown in Fig. 2 as a function of the scattering 
energy. 
For $E>2.33$ MeV, width of these two eigenvalues grow apart very fast. $E_1(E)$ 
(solid line) and $E_2(E)$ (dotted line) cross again for $E\simeq 3.2$ MeV. At this energy, 
$\mathit{\Gamma}_1$ and $\mathit{\Gamma}_2$ are different and, hence, the corresponding eigenfunctions are different as well. 

The upper part of Fig. 2 shows the phase shifts $\delta_{l_j}$ for 
${\rm p +} ^{15}{\rm F}$ elastic scattering as a function of the proton energy for $p_{1/2}$ 
(dashed-dotted line) and $p_{3/2}$ (dashed line) partial waves. In the partial wave 
$p_{1/2}$, the elastic scattering phase shift exhibits a jump by $2\pi$ at the EP with $J^{\pi}=1^-$. This unusual jump in the elastic scattering phase shift is an unmistakable and robust signal of a double-pole of the S-matrix (EP) which persists also in its neighborhood, as shall be discussed below. 

\begin{figure}[hbt]
\begin{center}
\includegraphics[width=7.5cm,angle=00]{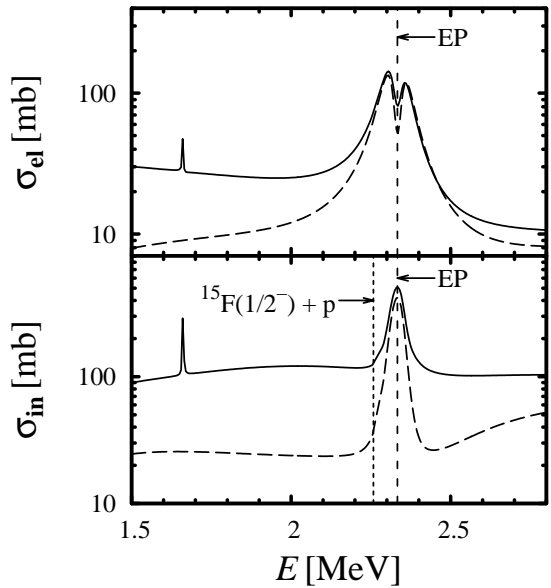}
\caption{Elastic and inelastic cross-sections in the reaction $^{15}{\rm F(p,p')}$ as 
a function of the proton energy $E$ at around the EP (the double-pole of the S-matrix) with $J^{\pi}=1^-$ for $1^-$ resonances only (dashed line) and for all resonances with $J\leq5$ (solid line).  For more details, see the description in the text.}
\label{fig3}
\end{center}
\end{figure}
Fig. 3 shows the elastic and inelastic cross sections for $^{15}{\rm F(p,p')}$ in the 
vicinity of an EP. The solid line represents a sum of different partial contributions of both parities 
with $J \le 5$ whereas the dashed line shows the resonance part of $1^-$ contribution in these cross sections. The cross sections are plotted as a function of center of mass scattering energy for 
$V_0^{(\rm cr)}=-1617.4$ MeV$\cdot$fm$^3$. The elastic cross section at the EP 
shows a characteristic double-hump shape \cite{Mul95} with asymmetric tails in 
energy. The  inelastic  cross section in this case exhibits a single peak. 
Both inelastic channels $[^{15}{\rm F}(5/2^+)\otimes {\rm p}_{l_j}]_{E'}^{1^-}$ 
and $[^{15}{\rm F}(1/2^-)\otimes {\rm p}_{l_j}]_{E'}^{1^-}$ are opened at the EP. 
Substantial background contribution to both cross sections comes from broad resonances, mainly $0^+$ and $2^+$. A sharp peak at $E\simeq 1.65$ MeV corresponds to an ordinary resonance $2^-$. 

The above discussion of the double-poles of the S-matrix (EPs) and their manifestation in the many-body scattering continuum concerns $1^-$ states. The same analysis for  $J^{\pi}=0^+, 2^+$ states of $^{16}$Ne gives qualitatively similar results. Also in these two cases, the number of EPs is maximal but only a fraction of them appears in the relevant range of $E$ and $V_0$ values.

\subsection{Behavior of scattering wave functions in the vicinity of the exceptional point}

A true crossing of two resonant states is accidental and,
\begin{figure}[hbt]
\begin{center}
\includegraphics[width=8cm,angle=00]{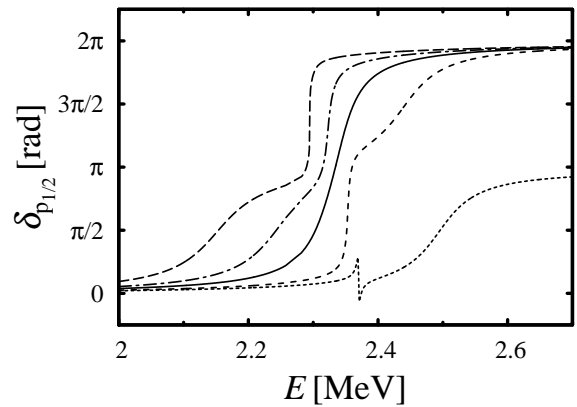}
\caption{The elastic scattering phase shifts $\delta_{p_{1/2}}$ for  ${\rm p +} ^{15}{\rm F}$ 
reaction in $1^-$ partial waves at around the EP (the double-pole of the S-matrix) with $J^{\pi}=1^-$ at 
$V_0^{(\rm cr)}=-1617.4$ MeV$\cdot$fm$^3$ (solid line). Different curves correspond to different strength $V_0$ of the continuum coupling: $V_0$=-1800 MeV$\cdot$fm$^3$ (long-dashed line), -1700 MeV$\cdot$fm$^3$ (dashed-dotted line), -1500 MeV$\cdot$fm$^3$ (short-dashed line) and -1430 MeV$\cdot$fm$^3$ (dotted line).} 
\label{fig4}
\end{center}
\end{figure}
 hence, improbable in nuclear scattering experimentation. In this section, we will investigate the behavior of scattering states  in the vicinity of an EP (the double-pole of the S-matrix) as the observation of such a situation is more plausible. 
 
Fig. 4 exhibits the phase shifts $\delta_{l_j}$ for  ${\rm p +} ^{15}{\rm F}$ elastic scattering as a function of the proton energy for various values of the strength  $V_0$ ($V_0$=-1800 MeV$\cdot$fm$^3$ (long-dashed line), -1700 MeV$\cdot$fm$^3$ (dashed-dotted line), -1617.4 MeV$\cdot$fm$^3$ (solid line), -1500 MeV$\cdot$fm$^3$ (short-dashed line) and -1430 MeV$\cdot$fm$^3$ (dotted line)) of the residual coupling $H_{QP}=H_{PQ}=V_0\delta(r_1-r_2)$ between ${\cal Q}$ and ${\cal P}$ subspaces. The characteristic change by a $2\pi$ of the elastic phase shift is seen in a broad interval -1800 MeV$\cdot$fm$^3$ $\leq V_0 \leq$ -1500 MeV$\cdot$fm$^3$ of the continuum coupling strength.

\begin{figure}[hbt]
\begin{center}
\includegraphics[width=6cm,angle=00]{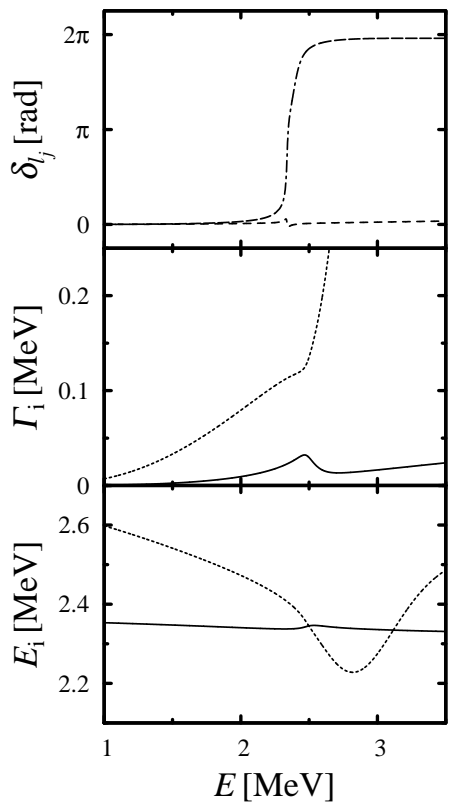}
\caption{The same as in Fig. \ref{fig2} but in the subcritical regime of coupling ($V_0=-1560$ MeV$\cdot$fm$^3$). For more details, see the caption of Fig. \ref{fig2} and the description in the text.}
\label{fig5}
\end{center}
\end{figure}
%
\begin{figure}[hbt]
\begin{center}
\includegraphics[width=6cm,angle=00]{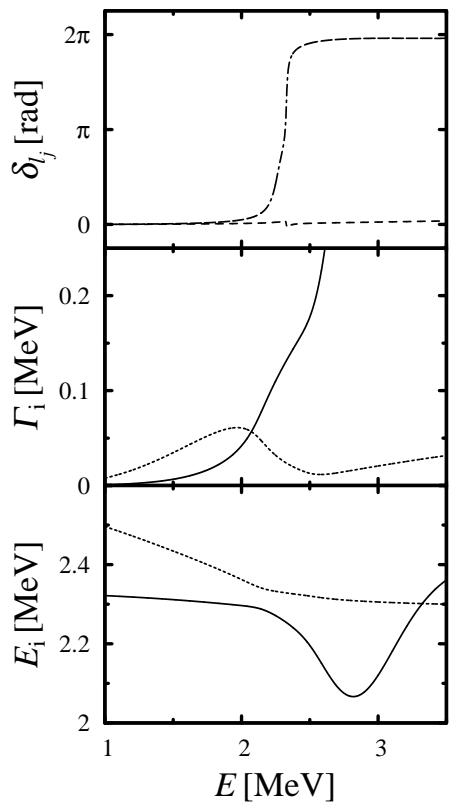}
\caption{The same as in Fig. \ref{fig2} but in the overcritical regime of coupling ($V_0=-1680$ MeV$\cdot$fm$^3$). For more details, see the caption of Fig. \ref{fig2} and the description in the text.}
\label{fig6}
\end{center}
\end{figure}
Fig. 5 and 6 show energies $E_i$ and widths $\mathit{\Gamma}_i$ of $1^-_1$ and $1^-_2$ eigenvalues 
as a function of the scattering energy for two values of $V_0$: -1560 MeV$\cdot$fm$^3$ (Fig. 5) and -1680 MeV$\cdot$fm$^3$ (Fig. 6). 

The case shown in Fig. 5 corresponds to a subcritical coupling where two resonances cross freely in energy and repel in width \cite{Phi00}. In this regime, the scattering energy $E$ corresponding to the closest approach of $1^-$ eigenvalues in the complex plane ($E\simeq 2.47$ MeV) is higher than the scattering energy corresponding to the EP at a critical coupling $V_0^{(\rm cr)}$=-1617.4 MeV$\cdot$fm$^3$.
Nevertheless, the elastic scattering phase shift shows the jump by $2\pi$ at the position of the EP and not at the point of the closest approach of eigenvalues.

Fig. 6 shows the situation corresponding to an overcritical coupling where two resonances exhibit level repulsion in energy and a free crossing of their widths \cite{Phi00}.  In this case, the point of the closest approach of $1^-$ eigenvalues in the complex plane is found at the scattering energy ($E=2.13$ MeV) which is lower than than the corresponding energy for the EP. Again, the  elastic scattering phase shift shows the jump by $2\pi$ at the position of the double-pole. 

From these two examples, one can see that the characteristic jump by  $2\pi$ of the elastic scattering phase shift  remains a robust signature of the EP in all close-to-critical regimes of the coupling to the continuum: the subcritical coupling ($|V_0|<|V_0^{(\rm cr)}|$), the critical coupling ($|V_0|=|V_0^{(\rm cr)}|$), and the overcritical coupling ($|V_0|>|V_0^{(\rm cr)}|$), where real and/or imaginary parts of two eigenvalues coincide. 

\begin{figure}[hbt]
\begin{center}
\includegraphics[width=7.5cm,angle=00]{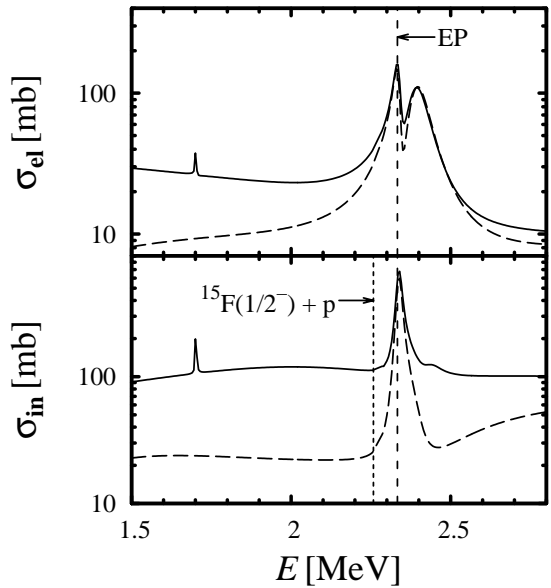}
\caption{The same as in Fig. \ref{fig3} but in the subcritical regime of coupling ($V_0=-1560$ MeV$\cdot$fm$^3$). For more details, see the caption of Fig. \ref{fig2} and the description in the text.}
\label{fig7}
\end{center}
\end{figure}
%
\begin{figure}[hbt]
\begin{center}
\includegraphics[width=7.5cm,angle=00]{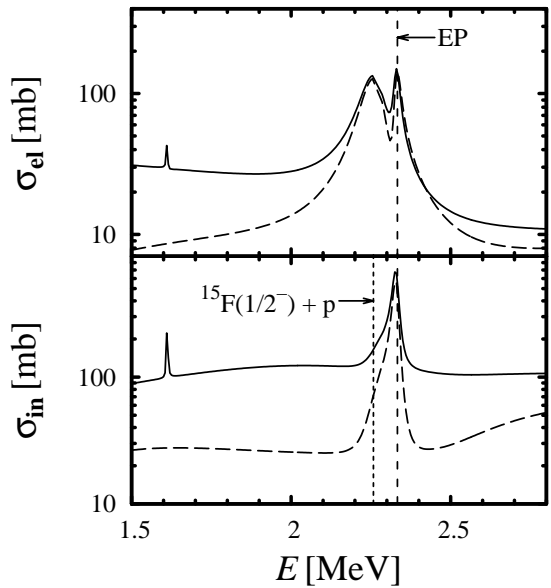}
\caption{The same as in Fig. \ref{fig3} but in the overcritical regime of coupling ($V_0=-1680$ MeV$\cdot$fm$^3$). For more details, see the caption of Fig. \ref{fig2} and the description in the text.}
\label{fig8}
\end{center}
\end{figure}
Next two figures show the elastic and inelastic cross sections for $^{15}{\rm F(p,p')}$ in the vicinity of the EP with $J^{\pi}=1^-$ in the subcritical (Fig. 7) and overcritical (Fig. 8) regimes of the continuum coupling. The curves shown by solid lines in Figs. 7,8 represent a sum of different partial contributions of both parities with $J \le 5$. The  curves shown by dashed lines exhibit the resonance part of  $1^-$ contribution in these cross sections. The qualitative features of the cross sections for the subcritical ($V_0=-1560$ MeV$\cdot$fm$^3$) and overcritical ($V_0=-1680$ MeV$\cdot$fm$^3$) couplings remain same as for the critical coupling (see Fig. 3). In both cases, one see a double-hump shape in the elastic cross sections and a single-hump shape in the inelastic cross section. One observes also a strong asymmetry in widths and heights of two peaks and a small shift of the position of the interference minimum in between the two peaks with respect to the energy  which the EP is found for a critical coupling.

\subsection{Entangled eigenstates of the effective Hamiltonian}

Complex and biorthogonal eigenstates of the effective non-Hermitian Hamiltonian 
provide a  convenient basis in which the resonant part of the scattering function can 
be expressed. These eigenstates are obtained by an orthogonal and, in general, non-unitary 
transformation of SM eigenstates \cite{Oko03} which is a consequence of their mixing
via coupling to common decay channels. The same coupling is responsible for the entanglement of two eigenstates involved in building of an EP, as illustrated in Fig. 9 on the example of spectroscopic factors. 

\begin{figure}[hbt]
\begin{center}
\includegraphics[width=6cm,angle=00]{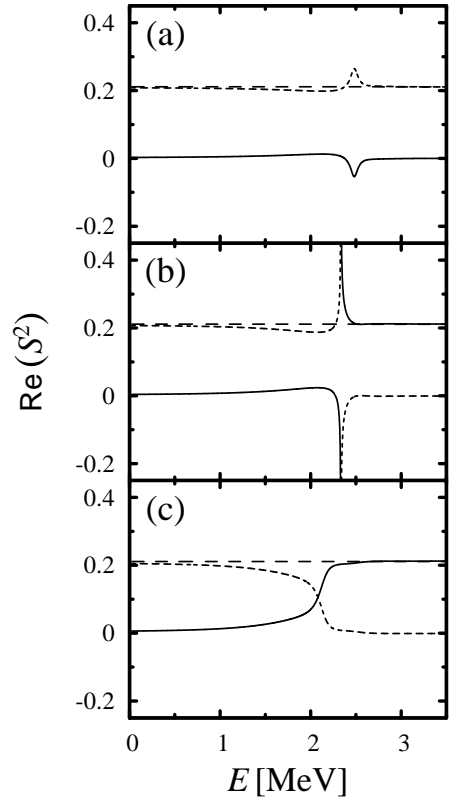}
\caption{$p_{1/2}$-spectroscopic factor 
$\langle^{16}{\rm Ne}(1_n^-)|[^{15}{\rm F}(1/2_1^+)\otimes p(0p_{1/2})]^{1^-}\rangle$ 
for $1_1^-$ and $1_2^-$ eigenvalues of the effective Hamiltonian at around the double-point of the 
S-matrix. For more details, see the discussion in the text.}
\label{fig9}
\end{center}
\end{figure}
Fig. 9 exhibits the real part of the spectroscopic factor Re($S^2$)=Re$\left(\langle^{16}{\rm Ne}(1_n^-)|[^{15}{\rm F}(1/2_1^+)\otimes p(0p_{1/2})]^{1^-}\rangle^2\right)$ in $^{16}$Ne
in three regimes of continuum coupling: (a) the subcritical regime ($V_0=-1560$ MeV$\cdot$fm$^3$), (b) the critical regime ($V_0^{(\rm cr)}=-1617.4$ MeV$\cdot$fm$^3$), and (c) the overcritical regime ($V_0=-1680$ MeV$\cdot$fm$^3$). The solid (short-dashed) lines show the spectroscopic factors for  $\Phi(1^-_1)(E)$ ($\Phi(1^-_2)(E)$) eigenvalues of the effective Hamiltonian ${\cal H}_{QQ}(E)$ as a function of the scattering energy $E$. For a critical coupling (plot (b)), the spectroscopic factors for  
$\Phi(1^-_1)$ and $\Phi(1^-_2)$ wavefunctions diverge at the EP (the double-pole of the S-matrix) but their sum (long-dashed line in Fig. 9) remains finite and constant over a whole region of scattering energies surrounding the EP.  In that sense, $\Phi(1^-_1)$ and $\Phi(1^-_2)$ resonance wavefunctions form an inseparable doublet of eigenfunctions with entangled spectroscopic factors.  
This entanglement is a direct consequence of the energy dependence of coefficients $b_{\alpha i}$:
\begin{eqnarray}
|\Phi_{\alpha}\rangle=\sum_i b_{\alpha i}(E) |\psi_i\rangle \ ,
\nonumber
\end{eqnarray}
in a decomposition of ${\cal H}_{QQ}(E)$ eigenstates in the basis of SM eigenstates. 

One may notice that the energy dependence of Re($S^2$) in the vicinity of the double-pole for $1^-_1$ and $1^-_2$ eigenstates is quite different in all three regimes of the continuum coupling. In particular, in the overcritical regime of coupling, an EP yields entangled states in a broad range of scattering energies. The strongest entanglement is found at the scattering energy which corresponds to the point of the closest approach of eigenvalues in the complex plane for all regimes of coupling. Obviously, the entanglement of resonance eigenfunctions in the vicinity of an EP is a generic phenomenon in open quantum systems which is manifested in matrix elements and expectation values for any operator which does not commute with the Hamiltonian.

\section{Conclusions}
In conclusion, we have shown in SMEC studies of the one-nucleon continuum that EPs  
exist for realistic values of the continuum coupling strength. In the studied case of  
$^{16}$Ne, few of those EPs appear at sufficiently low excitation energies to be seen in the excitation function as individual peaks associated with a jump by $2\pi$ of the elastic scattering phase shift. The occurrence of  an EP  leaves also characteristic imprints in its neighborhood, i.e. for avoided crossing of resonances. In all close-to-critical regimes of the continuum coupling where real and/or imaginary parts of the two eigenvalues coincide, one finds qualitatively similar features of the elastic scattering phase shift and the elastic cross-section as found for the critical coupling at around the EP (the double-pole of the S-matrix). This gives a real chance that EPs or their traces may actually be searched for experimentally in the atomic nucleus. The well-known case of $2^+$ doublet in $^8$Be, where resonance energies and widths are $16623\pm 3$ keV, $107\pm 0.5$ keV and $16925\pm 3$ keV, $74.4\pm 0.4$ keV, respectively \cite{Hin78}, nearly satisfies the resonance conditions in the close-to-critical regime of couplings. Various situations in this regime have been studied experimentally in the microwave cavity \cite{Phi00}.

Avoided crossing of two resonances with the same quantum numbers provide the valuable information about the configuration mixing in open quantum systems. As the formation of any EP in the scattering continuum depends on a subtle interplay between internal Hamiltonian ($H_{QQ}$) and the coupling to the external environment of decay channels, its finding provides a stringent test of an effective nucleon-nucleon interaction and the configuration mixing in the open quantum system regime. Such tests are crucial for a quantitative description of atomic nuclei in the vicinity of drip lines.

\vspace{0.2cm}
We wish to thank W. Nazarewicz for stimulating discussions and suggestions.

\end{document}